\documentclass[runningheads]{llncs}
\usepackage{graphicx}
\usepackage{bm}
\usepackage{amssymb,amsmath}
\usepackage{cite}
\usepackage{color}
\usepackage{soul}
\usepackage[noframe]{showframe}
\usepackage{framed}

\newcommand{\ak}{\textcolor{black}}
\newcommand\Rey{\mathrm{Re}}

\usepackage{color}

\renewenvironment{shaded}{%
  \MakeFramed{\advance\hsize-\width \FrameRestore\FrameRestore}}%
 {\endMakeFramed}
\definecolor{shadecolor}{gray}{0.75}

\begin{document}
\title{Auto-Encoded Reservoir Computing  for Turbulence Learning\thanks{The authors acknowledge the support of TUM-IAS, funded by the German Excellence Initiative and the EU 7th Framework Programme (grant no. 291763) and PRACE for awarding access to ARIS at GRNET, Greece. L.M. also  acknowledges the RAEng Research Fellowship Scheme.}}
\titlerunning{AE-RC for Turbulence Learning}
%
\author{N.A.K. Doan\inst{1,2} \and
W. Polifke\inst{2} \and
L. Magri\inst{3,4,5,6}
}
\authorrunning{N.A.K. Doan et al.}
%
\institute{Department of Aerospace Engineering, Delft University of Technology, Netherlands \and
Department of Mechanical Engineering, Technical University of Munich, Germany \and
Department of Engineering, University of Cambridge, UK \and
(visiting) Institute for Advanced Study, Technical University of Munich, Germany
\and The Alan Turing Institute, London, UK \and Imperial College London, Aeronautics department, London, UK}
\maketitle              
\begin{abstract}
We present an Auto-Encoded Reservoir-Computing (AE-RC) approach to learn the dynamics of a 2D turbulent flow. The AE-RC consists of an Autoencoder, which discovers an efficient manifold representation of the flow state, and an Echo State Network, which learns the time evolution of the flow in the manifold. The AE-RC is able to both learn the time-accurate dynamics of the flow and predict its first-order statistical moments. The AE-RC approach opens up new possibilities for the spatio-temporal prediction of turbulence with machine learning.

\keywords{Echo State Network \and Autoencoder \and Turbulence.}
\end{abstract}

\section{Introduction}
The spatio-temporal prediction of turbulence is challenging because of the extreme sensitivity of chaotic flows to perturbations, the nonlinear interactions between turbulent structures of different scales, and the unpredictable nature of  energy/dissipation bursts. 
Despite these intricate characteristics of turbulence, many advances have been achieved in its understanding with, for example, the energy cascade concept that provides a statistical description of the energy transfer between different scales in turbulent flows  \cite{Kolmogorov1941a}. Additionally, the existence of coherent structures, such as vortices, which evolve in a deterministic way, provides a basis for understanding turbulence \cite{Yao2020}:  
Within the chaotic dynamics of turbulence, there exist identifiable patterns that can help us predict the evolution of turbulent flows.
To discover such patterns, recent works have relied on machine learning \cite{Brunton2020}. In particular, the dynamics of models of turbulent flows have been learned by recurrent neural networks (RNNs) such as the Long Short-Term Memory units \cite{Wan2018,Srinivasan2019} or a physics-informed reservoir computing (RC) approach, based on Echo State Networks (ESN) \cite{Doan2020a}. 
Because RNNs are generally limited to low-dimensional datasets due to the complexity of training, past studies have been restricted to fairly low-dimensional systems. To deal with high dimensional fluid mechanical systems, recent approaches based on convolutional neural networks (CNNs), and in particular Autoencoders (AE), have shown great potential in discovering coherent structures in turbulent flows and reducing the dimensionality of flows \cite{Brunton2020,Murata2019}, \ak{more efficiently than  linear reduced-order modelling approaches  (for a review of reduced-order models in fluids refer to \cite{Rowley2017}).}

In this paper, we propose the Auto-Encoded Reservoir Computing framework (AE-RC). This combines an ESN and an AE with the objective of learning the spatio-temporal dynamics of a 2D turbulent flow governed by the Navier-Stokes equations (the Kolmogorov flow). The flow is discussed in Section \ref{sec:kol_flow}. The AE-RC framework is presented in Section \ref{sec:method} and results are discussed in Section \ref{sec:results}. The final section summarizes the results and outlines avenues for future work.

\section{Turbulent flow}
\label{sec:kol_flow}
\begin{figure}[!ht]
    \centering
    \includegraphics[width=0.85\textwidth]{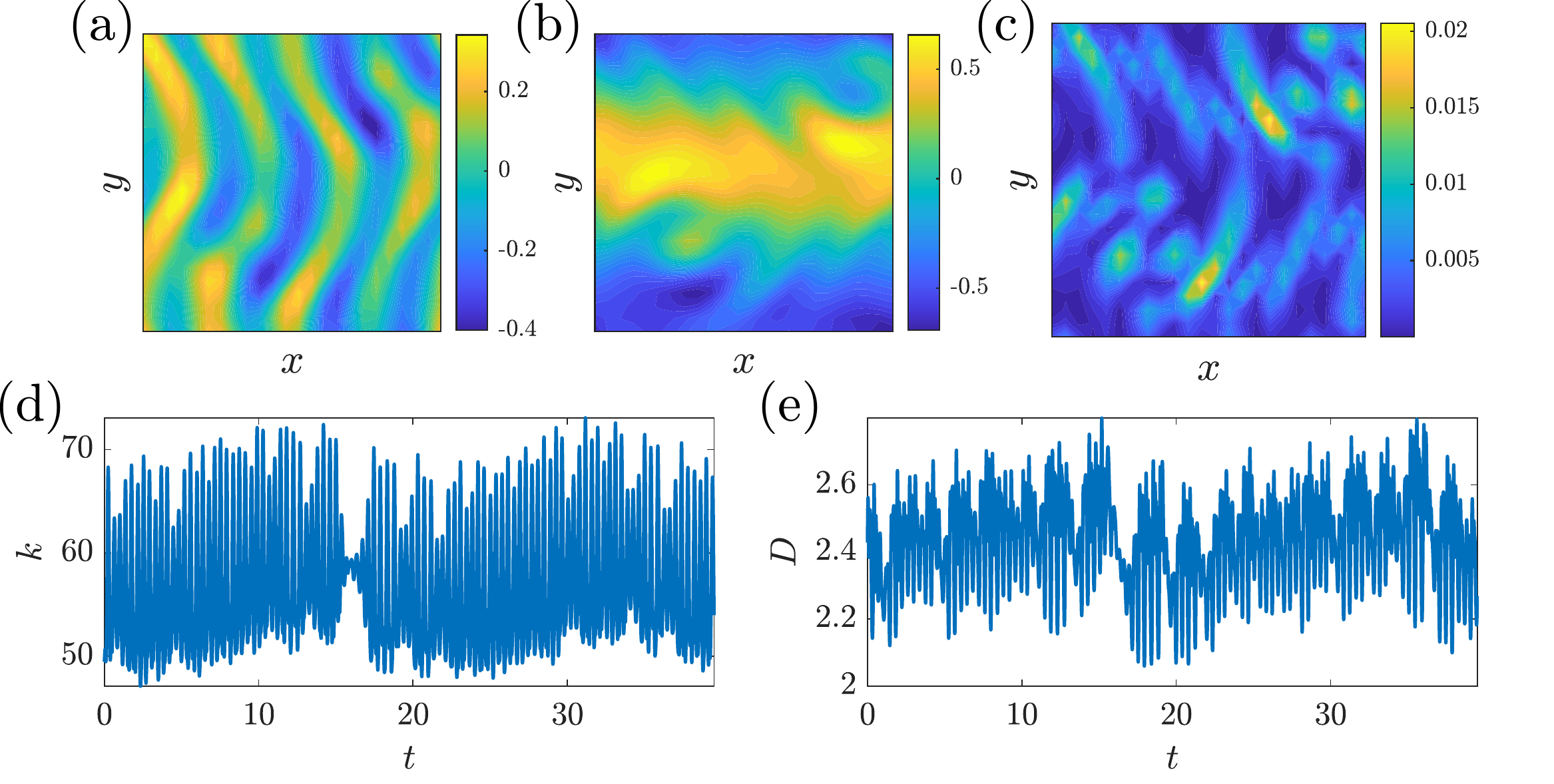}
    \caption{Isocontours of (a) $u$, (b) $v$ and (c) vorticity for the 2D turbulent flow at time $t=0$. (d) Time evolution of kinetic energy, $k$, and (e) dissipation, $D$.}
    \label{fig:Kol_flow_Re30}
\end{figure}

We investigate 2D turbulence governed by the incompressible Navier-Stokes equations 
\begin{align}
    \label{eq:kol_eq1}
    \nabla \cdot \bm{u} & = 0\\
    \label{eq:kol_eq2}
    \partial_t \bm{u} + \bm{u} \cdot \nabla {\bm{u}} & = - \nabla p + \frac{1}{\Rey} \Delta \bm{u} + \bm{f}
\end{align}
where $\bm{u}=(u,v)$ is the velocity field, $p$ is the pressure, $\Rey$ is the Reynolds number, and $\bm{f}$ is a harmonic volume force defined as $\bm{f} = (\sin(k_f y), 0)$ in cartesian coordinates. The Navier-Stokes equations are solved on a domain $\Omega \equiv [0, 2\pi] \times [0,2\pi]$ with periodic boundary conditions. (The solution of this problem is also known as the 2D Kolmogorov flow.)
The flow has a laminar solution $u=\Rey k_f^{-2} \sin (k_f y), v=0$, which is unstable for sufficiently large Reynolds numbers and wave numbers $k_f$ \cite{Platt1991}. Here, we take $k_f=4$ and $\Rey=30$ to guarantee the development of a turbulent solution \cite{Wan2018}.
The set of Eqs. (\ref{eq:kol_eq1}) and (\ref{eq:kol_eq2}) is solved on a uniform $N \times N$ grid, with $N=24$, using a pseudo-spectral code with explicit Euler in time \cite{Wan2018} with a timestep, $\Delta t=0.01$, to ensure numerical stability.
Snapshots of the velocity and vorticity, $\omega$, fields are shown in Fig. \ref{fig:Kol_flow_Re30}, in which the complexity and chaotic pattern of the turbulent flow can be observed.
Figures \ref{fig:Kol_flow_Re30}d and \ref{fig:Kol_flow_Re30}e show the time evolution of the kinetic energy, $k$, and dissipation, $D$, which are calculated as $k(\bm{u}) = (2\pi)^{-2} \int_\Omega \frac{1}{2} |\bm{u}|^2 d\Omega $ and $D(\bm{u}) = \Rey^{-1} (2\pi)^{-2} \int_\Omega |\nabla \bm{u}|^2 d\Omega $, respectively. The solution is turbulent. 

\section{Auto-Encoded Reservoir Computing}
\label{sec:method}
\begin{figure}
    \centering
    \includegraphics[width=0.95\textwidth]{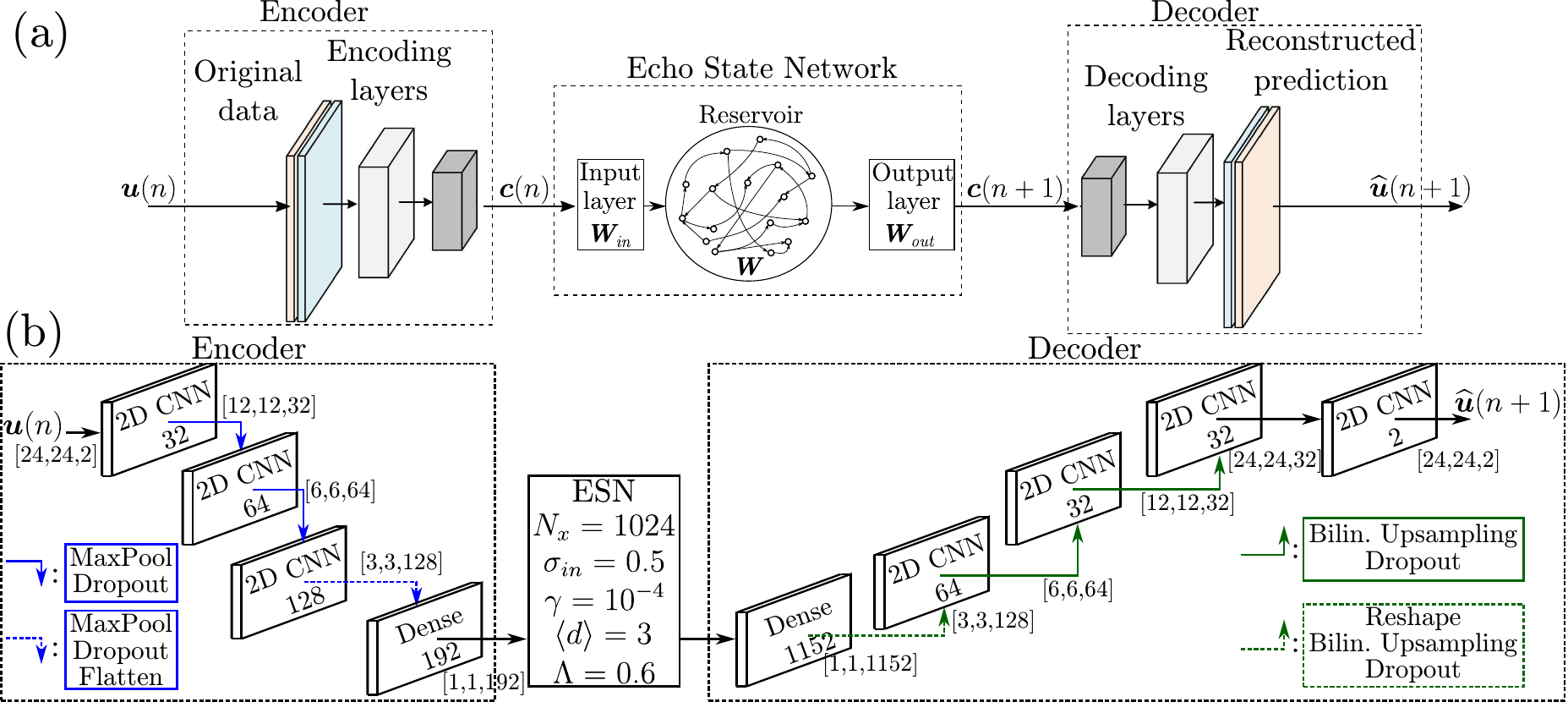}
    \caption{(a) Schematic and (b) details of the AE-RC. Number in boxes indicate the number of filters (for CNN) or neurons (for Dense layer). $[\cdot]$: dimension of the signal. $\tanh$ activation is used for all layers except the last layer in which a linear activation is used to reconstruct the velocity in the full range of real numbers. All MaxPool and upsampling layers have a window of $(2,2)$. All CNNs layers have a kernel of $(3,3)$.}
    \label{fig:AE_RC}
\end{figure}

The proposed Auto-Encoded Reservoir-Computing (AE-RC) framework is shown in Fig. \ref{fig:AE_RC}a. The AE-RC is composed of two parts: (i) an Autoencoder (AE), which is composed of an encoder and a decoder; and (ii) an echo state network, which is a form of reservoir computing \cite{Lukosevicius2009}. The role of the AE is to discover an efficient reduced-order representation of the original data, $\bm{u} \in \mathbb{R}^{N\times N \times 2=N_u}$. The encoder reduces the dimension of the data to a code, $\bm{c} \in \mathbb{R}^{N_c}$, where $N_c < N_u$, while the decoder reconstructs the data from the code, $\bm{c}$, by minimizing the error between the reconstructed solution, $\widehat{\bm{u}}$, and the data. Here, the AE consists of a series of CNNs, which identify patterns within images through kernel operations \cite{Goodfellow2016}. The details of the AE are shown in Fig. \ref{fig:AE_RC}b. On the downsampling side, the encoder is composed of multiple blocks of successive 2D CNNs, max pooling and dropout layers. Dropout layers prevent overfitting, while max pooling layers decrease the dimension of the input data. The dropout rate is 0.001 and was chosen during the training of the AE \ak{to have mean-squared errors of the same order of magnitude (and as small as possible) on both training and validation datasets.} \ak{(The dropout rate is rather small because the AE-RC has a small number of trainable weights with respect to the size of the dataset, which reduces the risk of overfitting).} 
After the last layer of the encoder, a dense feedforward neural network is used to combine the information from the previous layer and compress the data into the final code of dimension 192, compared to the original data of dimension $24 \times 24 \times 2=1152$. On the upsampling side, the architecture of the decoder mirrors that of the encoder, but the dimension of the code is progressively increased using bilinear upsampling layers to recover the original data~\cite{Murata2019}.

To learn the temporal dynamics of the reduced representation obtained with the AE, an Echo State Network (ESN) \cite{Lukosevicius2009} is employed as ESNs are accurate learners of chaotic dynamics and flows, e.g.,  \cite{Lukosevicius2009,Doan2020a}. The ESN receives the code as an input at a time $n$, $\bm{c}(n)$, and approximates the code at the subsequent time step, $\bm{c}(n+1)$, as an output. An ESN is composed of three parts:
(i) a randomized high dimensional dynamical system, called the reservoir, whose states of neurons (or units) at time $n$ are represented by a vector, $\bm{x}(n) \in \mathbb{R}^{N_x}$, $N_x$ being the number of neurons; 
(ii) an input matrix, $\bm{W}_{in}\in \mathbb{R}^{N_x \times N_c}$, and (iii) an output matrix, $\bm{W}_{out}\in \mathbb{R}^{N_y \times N_x}$ where $N_y$ is the dimension of the output of the ESN. The output of the ESN, $\widehat{\bm{y}}$, is a linear combination of the reservoir states, $\widehat{\bm{y}}(n)= \bm{W}_{out} \bm{x}(n)$. The evolution of the neurons' states is given by the discrete nonlinear law
\begin{equation}
    \bm{x}(n) = \tanh \left( \bm{W}_{in} \bm{c}(n) + \bm{W} \bm{x}(n-1) \right)
\end{equation}
Because the aim is to predict the dynamics of the reduced-order representation, the output of the ESN is the predicted subsequent state of the reduced-order representation, i.e.,  $\widehat{\bm{y}}(n)\approx\bm{c}(n+1)$. In the ESN approach, $\bm{W}_{in}$ and $\bm{W}$, are randomly initialized once and are not trained. Only $\bm{W}_{out}$ is trained. The sparse matrices $\bm{W}_{in}$ and $\bm{W}$ are constructed to satisfy the Echo State Property. Following \cite{Doan2020a}, $\bm{W}_{in}$ is generated such that each row of the matrix has only one randomly chosen nonzero element, which is independently taken from a uniform distribution in the interval $[-\sigma_{in}, \sigma_{in}]$. Matrix $\bm{W}$ is constructed with an average connectivity $\langle d \rangle$, and the non-zero elements are taken from a uniform distribution over the interval $[-1,1]$. All the coefficients of $\bm{W}$ are then multiplied by a constant coefficient for the largest absolute eigenvalue of $\bm{W}$, i.e. the spectral radius, to be equal to a value $\Lambda$, which is typically smaller than (or equal to) unity. The exact parameters of the ESN used here are provided in Fig. \ref{fig:AE_RC}b. The training procedure to train the AE-RC is provided in the grey box below.

\begin{shaded}
\vspace{-4pt}
\noindent \textbf{AE-RC TRAINING PROCEDURE}
\vspace{-6pt}
\begin{enumerate}
\item \textbf{Pre-train the AE} with the 2D velocity field as input/output. The reconstruction error, $E=\frac{1}{N_t} \sum_{n=1}^{N_t}|| \bm{u}(n) - \widehat{\bm{u}}(n)||^2$ where $N_t$ is the number of samples, is minimized. The AE  learns an appropriate reduced-order representation, $\bm{c}$, of $\bm{u}$.
\item \textbf{Compute the reduced representation}, $\bm{c}(n)$, of the original dataset, $\bm{u}(n)$, using the encoder part of the the pre-trained AE.
\item \textbf{Pre-train the ESN} using the dataset $\bm{c}(n)$ and ridge regression, $\bm{W}_{out} = \bm{Y} \bm{X}^T \left( \bm{X} \bm{X}^T + \gamma \bm{I}  \right)^{-1}$, where $\bm{Y}$ and $\bm{X}$ are the horizontal concatenation of the target data, $\bm{c}(n)$, and the associated ESN states $\bm{x}(n)$, respectively. $\gamma$ is the Tikhonov regularization factor \cite{Lukosevicius2009}.
\item \textbf{Train the combined AE-RC} for further fine-tuning. The AE-RC receives $\bm{u}(n)$ as an input and predicts $\widehat{\bm{u}}(n+1)$. The training minimizes $L=\frac{1}{N_t} \sum_{n=1}^{N_t}|| \bm{u}(n+1) - \widehat{\bm{u}}(n+1)||^2$, where $\widehat{\bm{u}}(n+1)$ is the prediction of the AE-RC at the next timestep, given an input $\bm{u}(n)$.
\end{enumerate}
\vspace{-13pt}
\end{shaded}
Steps 1 to 3 are used to obtain an initial AE-RC, which is the initial guess for the training of the entire AE-RC in Step 4. This accelerates the overall training of the AE-RC by taking advantage of the fast training of the ESN with ridge regression compared to a random initialization of the AE-RC. The ADAM optimizer \cite{Kingma2015} is used for Steps 1 and 4 with a learning rate of 0.0001.

\section{Results}
\label{sec:results}
\begin{figure}[!ht]
    \centering
    \includegraphics[width=0.85\textwidth]{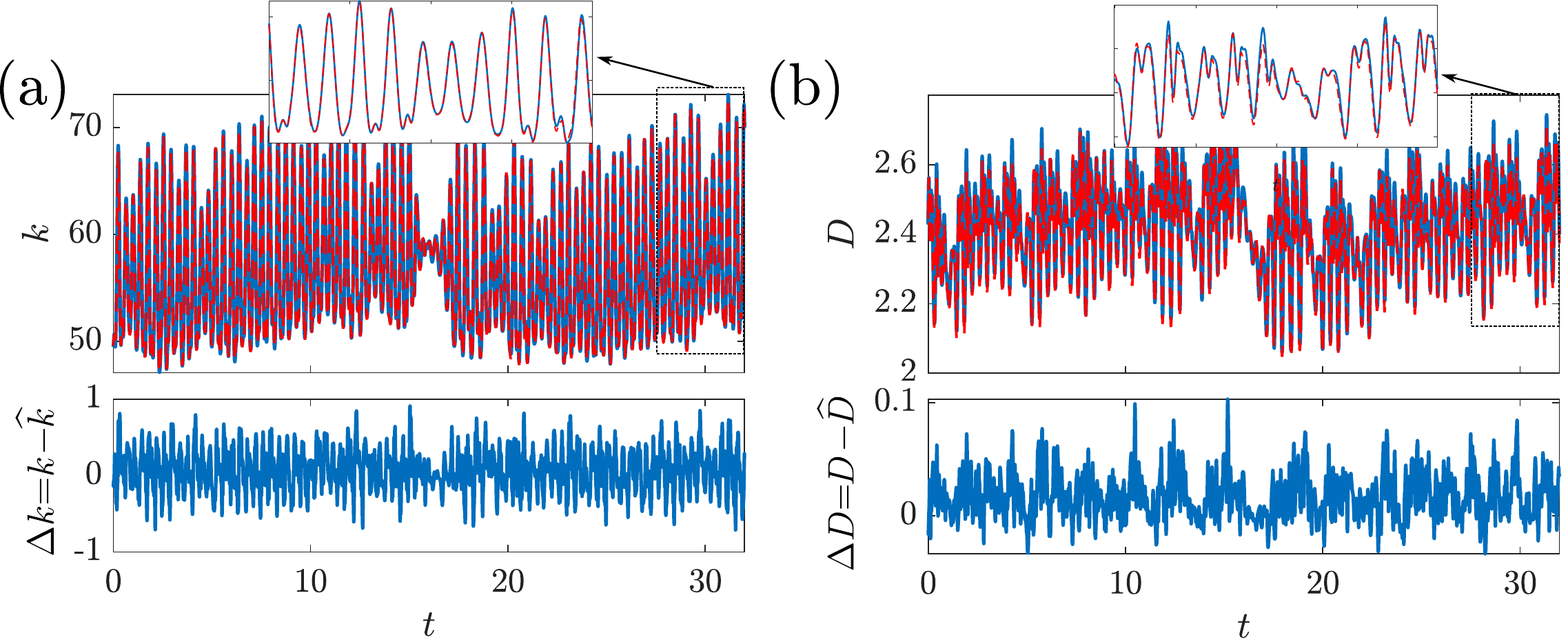}
    \caption{Time evolution of (a) $k$ and the error, $\Delta k$, and (b) $D$ and the error, $\Delta D$, for the 2D turbulent flow during the training stage. The AE-RC is ``teacher-forced'' (the input is provided by the training data). Blue lines: benchmark evolution. Dashed red lines: AE-RC.}
    \label{fig:AE_RC_Train}
\end{figure}

\begin{figure}[!ht]
    \centering
    \includegraphics[width=0.9\textwidth]{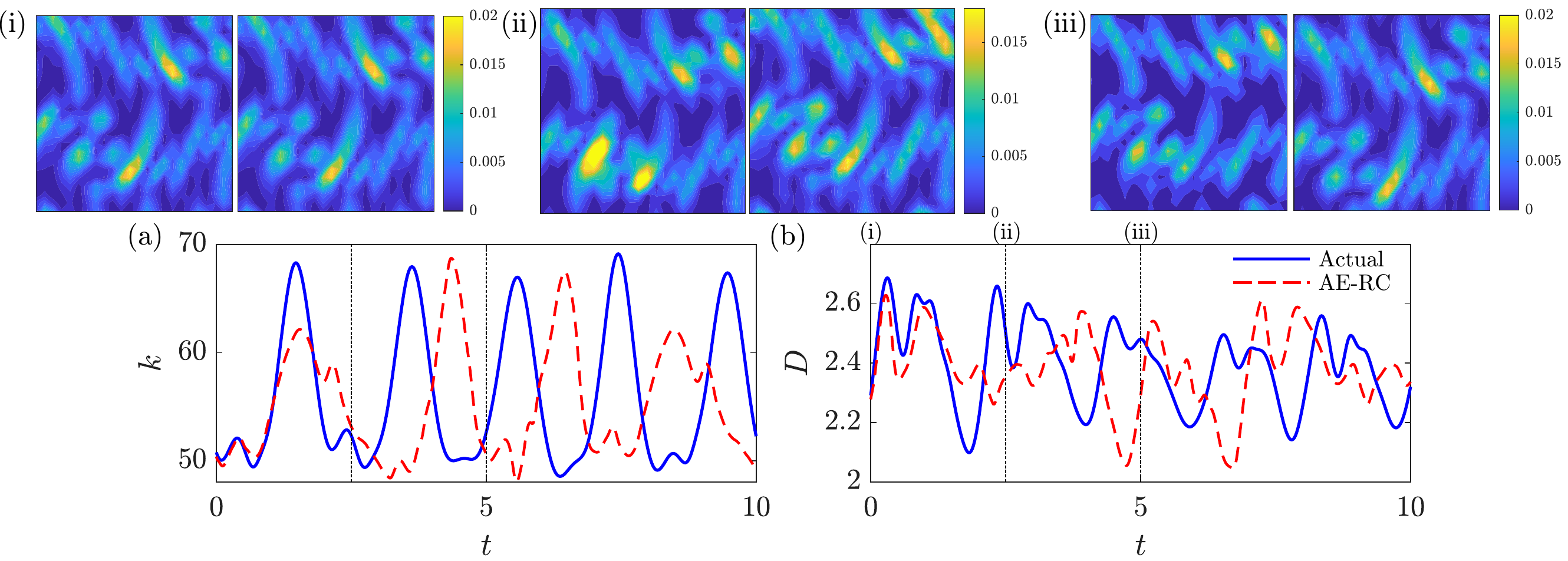}
    \caption{Time evolution of (a) $k$ and (b) $D$ for the 2D turbulent flow. Vertical dotted lines indicates the time-instants for the snapshots (i) to (iii) of vorticity. For each top panel: (left) benchmark evolution, (right) AE-RC prediction.}
    \label{fig:Dissip_predict}
\end{figure}

\begin{figure}[!ht]
    \centering
    \includegraphics[width=0.65\textwidth]{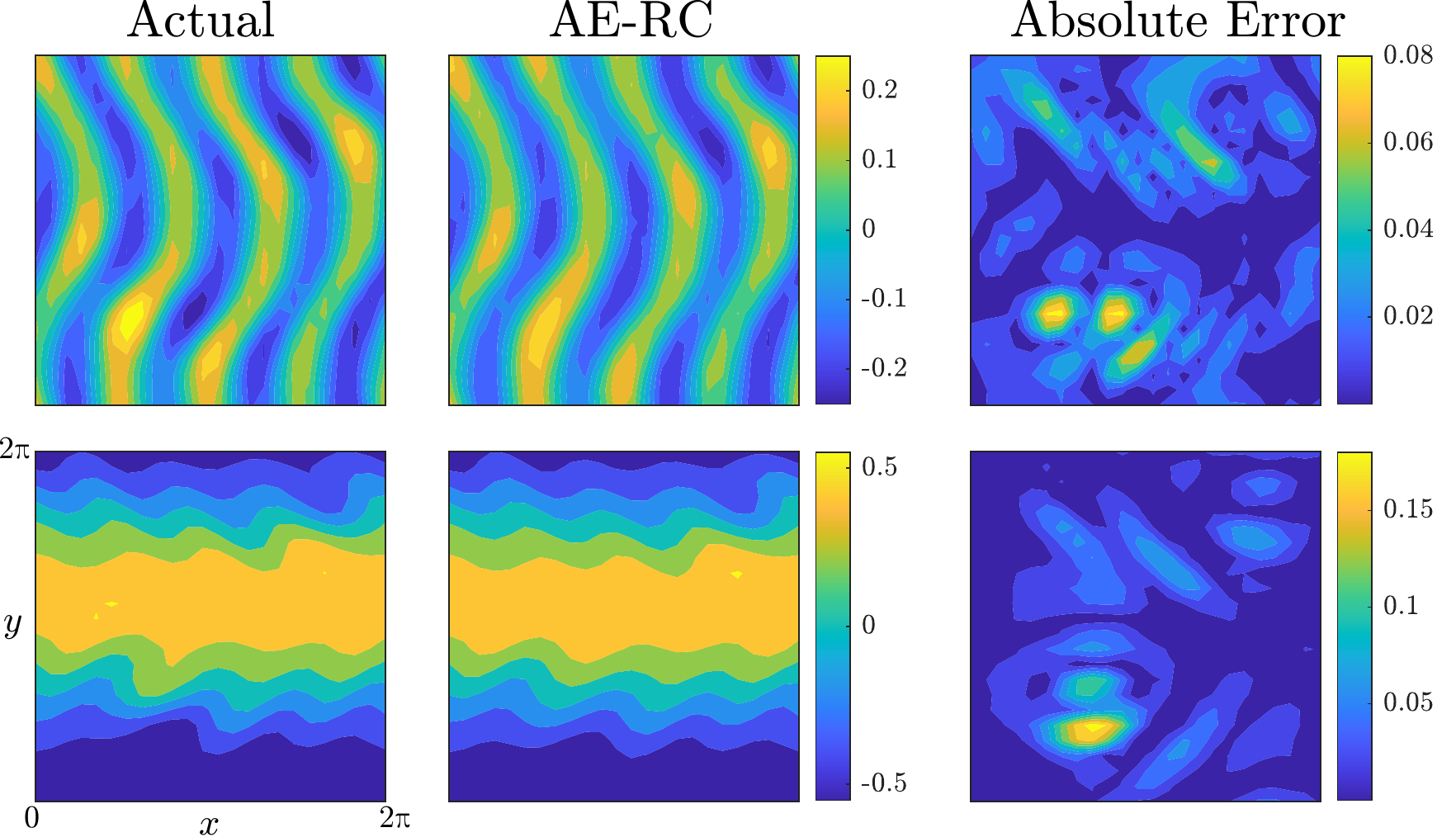}
    \caption{Prediction of time-averaged $u$ (top row) and $v$ (bottom row). Right column: absolute error between AE-RC prediction and benchmark solution.}
    \label{fig:Re30_VelMean}
\end{figure}

The AE-RC framework presented in Sec. \ref{sec:method} is applied to learning the dynamics of a 2D turbulent flow. The training dataset corresponds to the first 80\% of the time-evolution shown in Fig. \ref{fig:Kol_flow_Re30} and the last 20\% are used for validation.  The AE-RC receives the 2D velocity field at a given timestep, as the input, and predicts the velocity field at the next timestep, as the output. The predictions of $k$ and $D$ during training (quantities noted with $\hat{\cdot}$) are shown in Fig. \ref{fig:AE_RC_Train} \ak{with their  errors}. The AE-RC accurately reproduces the evolution in the training data. To assess the extrapolation capability, the output of the AE-RC is looped back as an input so that the AE-RC evolves autonomously.
The learned extrapolated time-series of $k$ and $D$ are shown in Fig. \ref{fig:Dissip_predict} (the insets of the vorticity fields are shown for different time instants). The AE-RC reproduces the spatio-temporal evolution of $k$ and $D$, which is in agreement with the physical evolution of the turbulent flow. \ak{The phase difference between the AE-RC solution and the benchmark solution} may be due to the spatio-temporally chaotic nature of the flow, in which small errors in the initial conditions are amplified exponentially in a short time. This is why, in turbulent flows, the statistics  are typically compared to assess the accuracy of a solution. Figure \ref{fig:Re30_VelMean} shows the time-averaged velocity profiles respectively, computed over the duration shown in Fig. \ref{fig:Dissip_predict}. \ak{Because the error is small (the average absolute error of $u$ and $v$ normalized by their respective maximum values is less than 6\% and 4\% respectively),} it is concluded that the AE-RC has learned the dynamics of the Kolmogorov flow also in a statistical sense (for the first moment). The standard deviations of the velocity profile were also computed and found to be of similar accuracy as those of time-averaged velocity (not shown here).

\section{Conclusions and future directions}
\label{sec:conclusion}

We propose the Auto-Encoded Reservoir-Computing framework (AE-RC) to learn the dynamics of high-dimensional turbulent flows, which are both spatially and temporally chaotic. This framework consists of an Autoencoder, which learns an efficient reduced-order representation of the spatial dynamics, and an Echo State Network, which learns the temporal dynamics of the reduced-order representation. With these two components, the AE-RC is able to learn both the instantaneous and average dynamics of the two-dimensional turbulent flow governed by the incompressible Navier-Stokes equations. This framework is being assessed on flow conditions that also exhibit bursts of kinetic energy. \ak{In future work, the effect of the code dimension on the accuracy of the AE-RC will be analysed. A comparative study of the AE-RC performance with respect to existing non-intrusive linear reduced-order models, such as the Proper Orthogonal Decomposition with Galerkin projection, and with respect to Long-Short-Term Memory units for the time prediction is scope for further research.}
%

\end{document}